# Photoionization current spectroscopy of individual silicon vacancies in silicon carbide

Kazuki Okajima[1], Tetsuri Nishikawa[1,2], Hiroshi Abe[3], Koichi Murata[4], Takeshi Ohshima[3,5], Hidekazu Tsuchida[4], Naoya Morioka[1,2,a], and Norikazu Mizuochi[1,2,6]

[1] Institute for Chemical Research (ICR), Kyoto University, Uji 611-0011, Japan
[2] Center for Spintronics Research Network, ICR, Kyoto University, Uji 611-0011, Japan
[3] National Institutes for Quantum Science and Technology (QST), Takasaki, 370-1292, Japan
[4] Central Research Institute of Electric Power Industry (CRIEPI), Yokosuka 240-0196, Japan
[5] Department of Materials Science, Tohoku University, Sendai, 980-8579, Japan
[6] International Center for Quantum-field Measurement Systems for Studies of the Universe and Particles (QUP), KEK, Tsukuba, 305-0801, Japan

[a] Author to whom correspondence should be addressed: morioka.naoya.8j@kyoto-u.ac.jp

**Abstract**
Defect charge-state dynamics are central to both spin–photon interfaces and photoelectrical spin readout. Despite the significance of silicon vacancies (V1/V2) in silicon carbide (4H-SiC) for both applications, their ionization behavior has remained unclear because their lack of optical blinking prevents conventional charge-state analysis. Here, we employ photocurrent spectroscopy of individual defects to measure the wavelength dependence of their excitation and ionization cross-sections. We reveal that V1 and V2 exhibit similar ionization cross-sections that increase toward shorter wavelengths, while carbon vacancies dominate the more steeply increasing background photocurrent. These results indicate that V2 and its surrounding environment appear more robust than V1 under resonant excitation. We also identify wavelength regimes that optimize defect-origin photocurrent for photoelectrical spin readout relative to background contributions, which differ between single-defect and ensemble measurements. Our results establish photocurrent spectroscopy as a powerful complement to optical methods, advancing the development of defect-based quantum devices.

The charge degree of freedom of point defects has emerged as a key factor in realizing solid-state quantum technologies. Stable charge states under optical irradiation are crucial for achieving bright emission and spectral stability in the spin–photon interface[1–3], whereas controlled charge cycling underpins efficient spin readout via photoelectrically detected magnetic resonance (PDMR)[4], enabling scalable and highly integrated quantum devices. These two applications impose opposite requirements in ionization: suppression or enhancement, yet both fundamentally rely on a quantitative understanding of ionization cross-sections and charge dynamics.

Negatively charged silicon vacancies ($S=3/2$)[5] in the 4H polytype of silicon carbide (4H-SiC), namely V1 and V2 at quasi-hexagonal and -cubic sites, respectively, and have attracted considerable attention as highly promising quantum defects[6–8]. Their excellent spin–photon properties, including long spin coherence[6,7,9,10] access to nuclear-spin memory[9,11], spectral stability[7,9,12,13], and recently demonstrated spin–photon entanglement[14] and nuclear-spin single-shot readout[15,16], make them strong candidates for quantum networking applications. Moreover, recent demonstrations of their integration into nanophotonic structures[9,17–19] and the enhanced spectral stability achieved by embedding defects in diode devices[20–22], which are compatible with wafer-scale semiconductor fabrication, highlight the technological maturity of SiC platform. In parallel, V2 centers have enabled highly sensitive photoelectrical spin readout via PDMR, from single-defect detection to large-ensemble defects including nuclear spins[23–25]. PDMR implementations likewise benefit from the device-integration advantages of a mature SiC electronics platform, which is available on large-scale wafers. Consequently, silicon vacancies in SiC are poised to play a central role in scalable quantum technologies.

Despite these advances, the charge stability and ionization efficiency of silicon vacancies under optical irradiation remain insufficiently understood. A major obstacle is the absence of a clear optical signature of charge-state transitions. Unlike nitrogen–vacancy centers in diamond, which exhibit photoinduced blinking[26,27], silicon vacancies rarely show detectable blinking under optical excitation[6,28], making their charge dynamics difficult to access optically. Recent low-temperature charge-resonance-check (CRC) techniques provide partial information on charge-state dynamics[21] including the charge fluctuations in the surrounding defect environment, yet the excitation-wavelength dependence of silicon-vacancy ionization has not been determined conclusively.

In PDMR experiments, in which photoionized carriers are measured for spin detection, the primary factor is the wavelength dependence of the excitation and ionization efficiencies of the relevant charge states of the target defect, as demonstrated for the nitrogen–vacancy center in diamond[29]. Further, an additional complication arises from background photocurrent. At room temperature, only the V2 center contributes to the spin-dependent signal. In contrast, the V1 center and other defects generate spin-insensitive photocurrent, increasing the background photocurrent and reducing the signal-to-noise ratio of V2 PDMR. Understanding the physical origin and spectral characteristics of both V2 and background photocurrents is therefore essential for optimizing PDMR measurements.

In this work, we address these challenges by directly detecting and imaging photocurrent from individual V1 and V2 centers. Direct charge measurements allow us to probe charge-cycling events, while the imaging enables us to disentangle defect-origin photocurrent from background contributions and to determine the excitation-wavelength dependence of both the ionization efficiency of silicon vacancies and the background defects. With these capabilities, we provide insight into the charge robustness of silicon vacancies and their surrounding environment for spin–photon interface applications, and identify the excitation-wavelength ranges that optimize PDMR for both single and ensemble V2 centers.

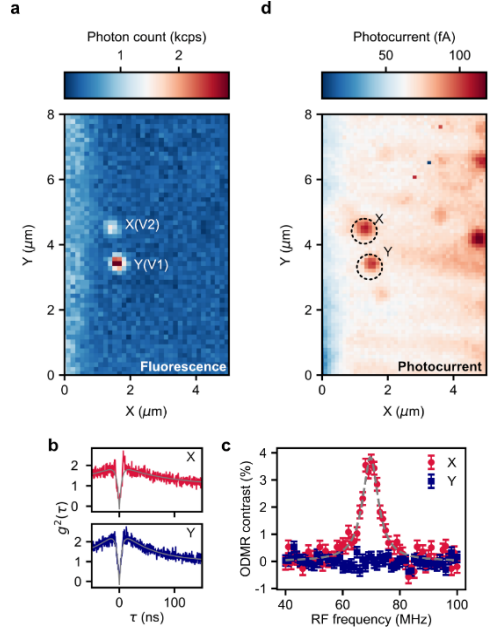

Figure 1. Device characterization. **a** Fluorescence image between the electrodes under 730 nm laser at 70 μW. **b** Autocorrelation functions $g^{(2)}(\tau)$ of defects X and Y. **c** ODMR spectra of defects X and Y. **d** The photocurrent image of the same region as **a** under 905 nm laser at 3 mW.

To investigate the photoionization of individual silicon vacancies, we performed photocurrent measurements on single defects using a metal–semiconductor–metal Schottky diode[25]. The starting material was an epitaxially grown p-type 4H-SiC film with a net acceptor concentration of $2 \times 10^{14}$ cm$^{-3}$, containing isolated silicon vacancies created by 2-MeV electron irradiation to a fluence of $8 \times 10^{13}$ cm$^{-2}$ and annealing at 600°C[25]. The device was fabricated on the surface with a pair of gold Schottky electrodes. By biasing above the flat-band condition, nearly ideal photocurrent collection is expected[30,31]. The electrode gap is approximately 7.5 μm, and the designed flat-band voltage is 10 V.

Photocurrent and optical measurements were performed using a home-built confocal laser-scanning microscope equipped with an oil-immersion objective lens (NA = 1.45) for excitation and detection. The system includes a transimpedance amplifier (NF SA-609F2, $10^{12}$ V/A), a 300-Hz low-pass filter, and a lock-in amplifier, enabling sensitive photocurrent detection. For optical measurements, a 730-nm laser excitation was used, and fluorescence above 900 nm was detected by two silicon single-photon-counting avalanche photodetectors in a Hanbury Brown and Twiss interferometer configuration. For photocurrent measurements, excitation wavelengths ranging from 789 to 940 nm were used. These lasers were delivered through the same single-mode fiber and achromatic collimator and focused onto the defects using the same objective lens as for the optical characterization. During the photocurrent measurement, the laser power was controlled by an acousto-optic modulator and modulated with an optical chopper wheel at 31 Hz for lock-in detection. The applied voltage was 10 V. The laser powers reported below are corrected for the objective lens's transmittance.

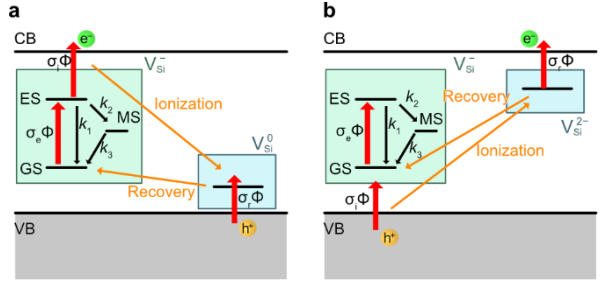

Figure 2. Photoionization dynamics of negatively charged silicon vacancies $V_{Si}^-$ in 4H-SiC. Two possible pathways are illustrated: **a** ionization into the neutral state $V_{Si}^0$ and **b** ionization into doubly negatively charged state $V_{Si}^{2-}$. The red arrows illustrate the laser-induced transitions: excitation, ionization, and charge recovery with their rates $\sigma_e \Phi$, $\sigma_i \Phi$, and $\sigma_r \Phi$, respectively, where $\sigma$ is an absorption cross-section for each process and $\Phi$ is a photon flux proportional to the laser power. The slanted orange arrows indicate the charge-state transitions accompanied by photocarrier generation in the conduction band (CB) and valence band (VB). The black arrows show spontaneous decay channels with associated rates ($k_1$, $k_2$, and $k_3$). Note that the specific charge state of the ionized configuration does not affect our analysis.

We first characterized the fabricated device. Figure 1a shows the fluorescence image between electrodes, where two isolated fluorescence spots were observed. We identified that one of them is a single V2 center (labelled X) from the autocorrelation function (Fig. 1b) and optically detected magnetic resonance (ODMR) measurements showing a characteristic zero-field resonance at 70 MHz (Fig. 1c). The other spot is also a single photon emitter (labelled Y) without any ODMR peaks in the resonance-frequency range of the V2 center (Fig. 1c). Given the defect creation conditions and the detection wavelength range, we attribute defect Y to a single V1 center. This assumption will be further justified later by studying the wavelength dependence of the excitation cross-section. In the same region, we characterized the photocurrent image (Fig. 1d). We observed isolated photocurrent spots at the same positions as defects X and Y, and confirmed a PDMR response at 70 MHz at the location of defect X (data not shown). These photocurrent signals therefore originate from the same defects identified optically.

To investigate the photoionization rate, we performed laser-power-dependent photocurrent measurements. We consider a four-level model consisting of the ground state (GS), excited state (ES), and metastable state (MS) of the negatively charged state, together with an additional level corresponding to the ionized state (Fig. 2). The ionized state is believed to be either neutral[23,25] or doubly negative[21]; although its exact charge has not been conclusively determined, this uncertainty does not affect our analysis. With this model, the dependence of the photocurrent ($I$) on the laser power ($P$) is given as:

$$I = \frac{\alpha P^2}{P + P_0}, \tag{1}$$

exhibiting a transition from a quadratic to linear power dependence. Here, $\alpha$ is the slope in the linear region and is proportional to the ionization cross-section, while $P_0$ is inversely proportional to the excitation cross-section and represents the power at which the quadratic and linear regimes cross over. Considering that silicon vacancies do not blink under photoexcitation[6,28], the charge recovery from the ionized state is expected to be much faster than the ionization from the negatively charged state. Under this assumption, the parameters $\alpha$ and $P_0$ are calculated to be[25]

$$\alpha = \frac{e\eta_c \gamma_i}{1 + k_2/k_3}, \qquad P_0^{-1} = \gamma_e \tau_{ES}\left(1 + \frac{k_2}{k_3} + \frac{\gamma_i}{\gamma_e}\right), \tag{2}$$

where $\gamma_e$ is the excitation rate from the GS to the ES, $\gamma_i$ is the photoionization rate from the ES to the ionized state, $e$ is the elementary charge, $\tau_{ES}$ is the ES lifetime (about 6 ns[28,32,33]), and $k_2/k_3$ is the ratio of the intersystem-crossing rates from ES to MS ($k_2$) and MS to GS ($k_3$). We used $k_2/k_3 \approx 7$ based on previous reports ranging 6–7.5[28,34], although the site effect, studied at cryogenic temperatures[35,36], has not been identified at room temperature. The photocurrent collection efficiency $\eta_c$ is assumed to be unity.

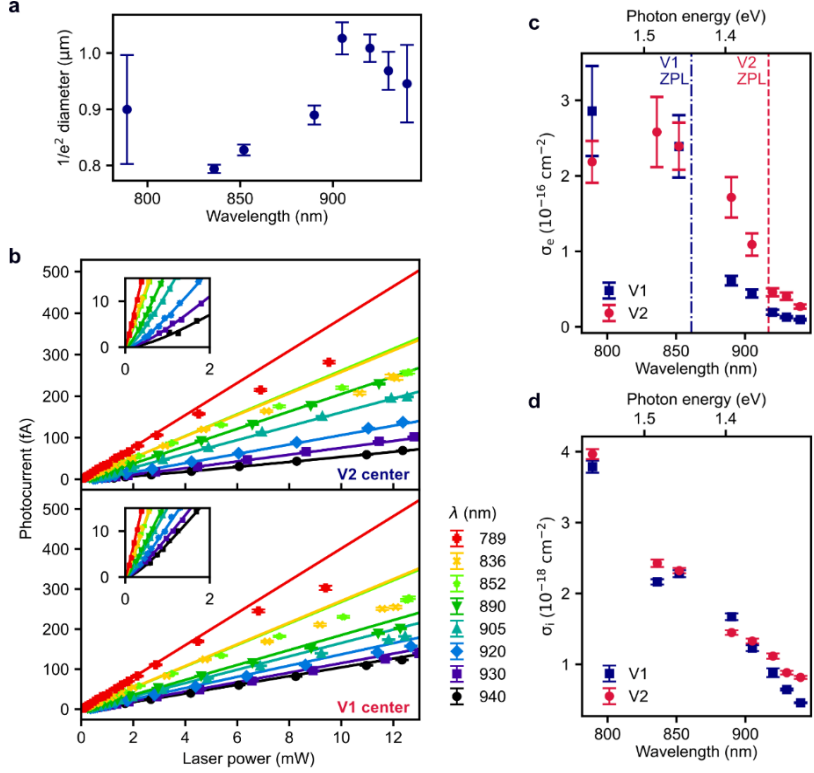

Figure 3. Laser-wavelength dependence of the photocurrent generation. **a** Estimated laser diameter from the spot-size analysis (defect Y). **b** Laser-power dependence of the photocurrent for the single V2 and V1 centers. The solid lines show fittings using Eq. (1). The insets show the zoom-in view of the low-power range. **c** Excitation-wavelength dependence of the excitation cross-sections for the V2 and V1 centers. The data point at 836 nm for V1 was excluded due to the excessively large error. **d** Excitation-wavelength dependence of the ionization cross-sections for the V2 and V1 centers. All error bars in this figure represent ±1 standard error.

Using Eq. (1) as a fitting function for the laser-power-dependent photocurrent, we estimated $\gamma_e$ and $\gamma_i$. These rates were then converted to excitation and ionization cross-sections, $\sigma_e$ and $\sigma_i$, respectively, taking into account the laser spot size and the conversion factor from the root-mean square value of the fundamental harmonic measured by the lock-in amplifier to the DC photocurrent[25]. The single-defect photocurrent was obtained by photocurrent imaging and two-dimensional Gaussian fitting, which also gives background information. The laser spot size was estimated from the imaging spot size of Y at the high-power limit[25] and is summarized in Fig. 3a. The spot size is independent of the laser wavelength; thus, we used the averaged value (0.9 μm for a $1/e^2$ diameter) in the analyses below.

Figure 3b shows the laser-power dependence of the photocurrent generated from the single V2 center (X) and the possible V1 center (Y), respectively. A quadratic dependence appears at longer wavelengths, whereas the crossover to the linear regime shifts to lower powers at shorter wavelengths, indicating an increase in the excitation cross-section. The wavelength dependence of the excitation cross-section for both defects is compared in Fig. 3c. An increase below each zero-phonon line (ZPL; V1: 862 nm, V2: 917 nm) is observed, which further supports the identification of defect Y as a V1 center.

We also extracted the wavelength-dependent ionization cross-section, as plotted in Fig. 3d. In contrast to the excitation cross-section, the ionization cross-sections for V1 and V2 are very similar across all wavelengths measured and increase monotonically with decreasing wavelength (increasing photon energy).

The extracted ionization cross-sections are consistent with the previously reported value ($6.4 \times 10^{-19}$ $cm^2$ at 905 nm) obtained from room-temperature photocurrent measurements[25]. However, they are approximately two orders of magnitude larger than the values under off-resonant excitation in recent cryogenic-temperature CRC-based optical measurements[21]. Unless the ionization dynamics differ significantly between room temperature and cryogenic temperatures, this discrepancy likely originates from ionization events accompanied by rapid charge-state recovery without a shift in resonance wavelength, which remain undetectable in CRC-based measurements. Also, a recent theoretical prediction, assuming the ionization to the doubly negative state[21], suggested an ionization rate about $5 \times 10^3$ times higher than the previous low-temperature experimental report under off-resonant excitation. Our measured ionization rates are substantially closer to the theoretical values, yet a discrepancy of roughly two orders of magnitude still remains. The origin of this remaining discrepancy is not yet fully understood. Possible mechanisms include polarization effects, alternative ionization pathways and/or ionized state, and differences in charge-state dynamics between room temperature and cryogenic conditions, such as phonon-involved processes. Clarifying these effects will require further investigation and represent an important direction for future studies on ionization dynamics.

Next, we focus on the background photocurrent. Figure 4a (left) shows its wavelength dependence at various laser powers, measured in the region surrounding the V2 center (X). The background current increases toward shorter wavelengths, indicating that the charge-state environment around the defects becomes less stable under higher-energy excitation. Notably, a steeper rise is observed below the wavelength corresponding to a photon energy of approximately 1.5 eV. This threshold is close to the ionization energy of carbon vacancies reported in electron paramagnetic resonance measurements under photoexcitation (1.47–1.48 eV)[37], which has been attributed to ionization from the singly to the doubly positive charge state. Moreover, considering the charge-transition levels of silicon vacancies and carbon vacancies[38], the stable observation of negatively-charged silicon vacancies implies the presence of carbon vacancies in the singly positive charge state. Carbon vacancies typically exist in as-grown epitaxial layers at densities of $10^{12}$–$10^{13}$ $cm^{-3}$ [39] and are generated by MeV electron irradiation at concentrations roughly an order of magnitude higher than silicon vacancies[40]. In our sample, their estimated density is approximately $10^{13}$ $cm^{-3}$ [25], corresponding to roughly five to fifteen carbon vacancies within the laser focus. These considerations indicate that carbon vacancies are the primary source of background photocurrent in our device. This interpretation is consistent with recent cryogenic measurements by Steidl et al.[21], which attributed the observed optical-resonance instability of V2 centers to charge fluctuations likely originating from nearby carbon vacancies.

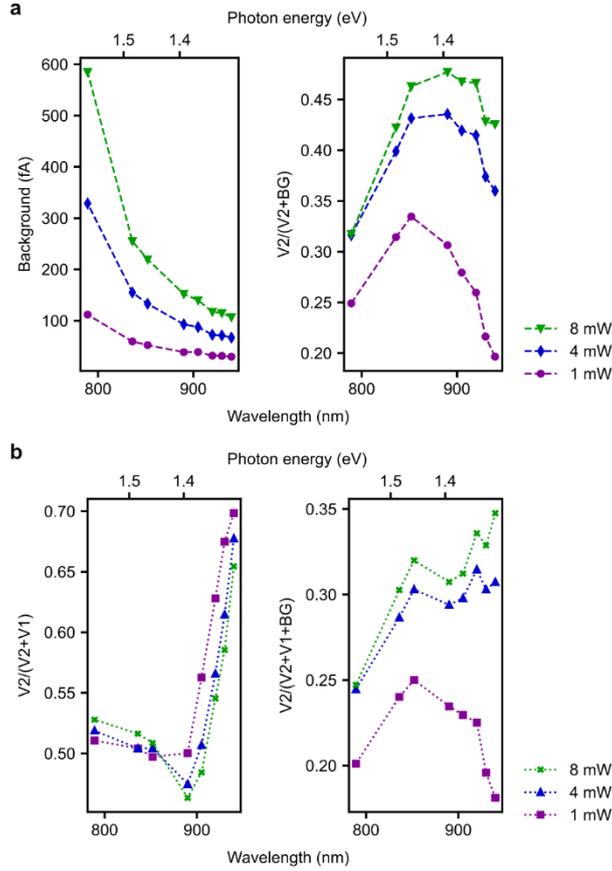

Figure 4. V2 center photocurrent fraction (PCF) analysis for PDMR applications. **a** Left: Excitation-wavelength dependence of the background measured around defect X; Right: The wavelength dependence of the PCF defined by V2/(V2+BG). **b** The wavelength dependence of the PCFs. Left: V2/(V2+V1); Right; V2/(V2+V1+BG). All PCF values are estimated by interpolating the experimental laser-power-dependent measurement results in Fig. 3b.

Based on the extracted ionization cross-sections (Fig. 3d), the V2 center is expected to be inherently less susceptible to photoionization than the V1 center under resonant excitation because of its longer ZPL wavelength. These findings are directly connected to higher charge stability, highlighting the advantage of V2 for spin–photon interface applications. In addition, the rapidly increasing background ionization at shorter wavelengths (Fig. 4a left) can limit spectral stability, particularly for V1, whereas it remains weaker near the V2 ZPL. Although a qualitative trend is anticipated from the general expectation that shorter-wavelength excitation enhances ionization, neither the ionization cross-sections of individual V1 and V2 centers nor the wavelength dependence of their surrounding charge environment had been quantitatively established prior to this work.

In addition, the pronounced ionization of background defects, primarily carbon vacancies, provides important insight into how repump or off-resonant excitation can modify the local charge environment, complementing earlier cryogenic observations of resonance fluctuations under resonant excitation.[21] These insights provide a quantitative basis for future studies on selecting repump and off-resonant excitation wavelengths that minimize unwanted ionization, thereby informing operational conditions for achieving charge-stable silicon-vacancy-based spin–photon interfaces.

We now consider the optimal excitation conditions for PDMR of the V2 center. The fundamental noise floor arises from the combined contributions of amplifier noise and photocurrent shot noise. At the same time, the laser power fluctuations add a further noise component, which has previously been suggested as the dominant noise source[25]. Because both contributions scale with the total photocurrent, through its square root for shot noise and linearly for laser-induced noise, reducing the background photocurrent is essential for improving the achievable sensitivity.

To quantify the relative contribution of the defect signal, we define the V2 photocurrent fraction (PCF) as V2/(V2+BG) (Fig. 4a right), which provides a photocurrent-derived measure of the single-defect PDMR performance. At high laser powers (8 mW), which are important for PDMR, the optimal excitation wavelength appears in the range of 852–920 nm, owing to the suppression of background-defect ionization. However, the PCF in the present sample is below 0.5, indicating that more than half of the photocurrent originates from the background defects rather than from the V2 center. This suggests that the presence of background photocurrents, primarily from carbon vacancies, increases the noise budget.

Because carbon vacancies are thermally more stable than silicon vacancies[41,42], they are difficult to remove after V2 formation. Nevertheless, carbon-vacancy densities can be reduced below $10^{11}$ cm$^{-3}$ [43–45] by annealing under carbon-interstitial-rich conditions, which can be prepared using established techniques such as carbon-ion implantation[43] or thermal oxidation[44,45]. These processes serve as effective pretreatments prior to silicon-vacancy creation, benefiting both PDMR and spin–photon interface applications. Furthermore, using lower electron fluence suppresses the absolute density of carbon vacancies, thereby reducing background photocurrent. Alternatively, irradiation schemes dominated by nuclear stopping, such as proton or light-ion implantation, can improve the silicon vacancy/carbon vacancy ratio (e.g., 1:1.5–1.7 reported by proton[46]). Although optimization of the electron fluence, implantation parameters, defect-generation yield, and annealing temperature is required, these approaches represent promising routes to suppress background photocurrent and enhance single-defect PDMR sensitivity.

While the optimization strategies discussed above directly address single-defect PDMR, the situation differs in ensemble systems. A primary distinction is that photocurrent from V1 centers also contributes to the total spin-insensitive signal. Recently, partially selective generation of V2 centers over V1 centers has been demonstrated using focused-ion-beam irradiation from the *m*-face rather than the conventional *c*-face.[47] Although this method can potentially contribute to the improvement of the spin contrast of ensemble V2 PDMR, *m*-face wafers are currently not scalable for electronic integration. A more practical strategy, motivated by our findings, is to suppress the photoionization of V1 centers.

Although ensemble measurements cannot separate V1 and background contributions, our single-defect data allow us to infer their expected ensemble behavior. As a performance indicator of ensemble PDMR based on photocurrent data, we consider V2/(V2+V1) and V2/(V2+V1+BG) as PCFs (Fig. 4b). These metrics assume that the densities of V1 and V2 centers are comparable and that the background defect density scales similarly to that of V2 centers with irradiation dose.

A higher PCF is obtained at longer wavelengths under anti-Stokes conditions for both V1 and V2 centers. Although the ionization cross-sections of V1 and V2 are similar, the inefficient excitation of V1 at wavelengths far longer than its ZPL suppresses its photocurrent contribution. While an ODMR study reported improved contrast by suppressing V1 excitation at wavelengths above 850 nm[10], our photocurrent measurements show that even longer wavelengths, above 900 nm, are required to suppress V1-origin photocurrent (Fig. 4b, left). This difference arises because high laser powers typically used for PDMR can still populate the V1 excited state even at wavelengths slightly above its ZPL, enabling ionization and photocurrent generation. At the same time, the photocurrent from the background defect is likewise reduced at longer wavelengths, improving the PCF (Fig. 4b right). Consequently, within the wavelength range examined up to 940 nm, longer-wavelength excitation yields higher PCF and is therefore advantageous for ensemble PDMR.

Beyond V1, V2, and carbon vacancies, additional defects may contribute to the background current in ensemble-defect PDMR. Photocurrent-only features visible in our maps (Fig. 1d and Ref. 25) likely originate from non-emissive or near-infrared emitters outside the detection range and generate photocurrents comparable to silicon vacancies. Their characterization lies beyond the scope of this work, but they represent an important direction for improving ensemble PDMR.

In summary, photocurrent spectroscopy of individual silicon vacancies reveals that the photoionization cross-sections of V1 and V2 are comparable and increase toward shorter wavelengths. We also find that carbon vacancies dominate the background photocurrent, affecting the charge environment. These results show that V2 is intrinsically more robust against ionization under resonant excitation than V1 and provide quantitative information for optimizing repump conditions in spin–photon interface applications. For PDMR, we identify an optimal excitation wavelength range of 852–920 nm for single V2 centers and a longer wavelength (~940 nm in the range studied) for ensembles, and show that reducing carbon-vacancy density can further improve spin-readout SNR. Our technique is compatible with recently developed defects-in-diode structures[20–22] to stabilize and tune optical transitions, and the combination of electrical and optical probing would enable access to charge-dynamic processes that remain invisible in purely optical methods. This work deepens the understanding of charge stability in spin–photon interface applications and provides quantitative guidelines for improving PDMR efficiency, thereby advancing scalable optical and electrical quantum-device platforms in silicon carbide.

**Acknowledgement**

Recently, we became aware of the wavelength-dependent PDMR study on ensemble silicon vacancies in high-purity semi-insulating 4H-SiC[48]. This work was supported by JSPS KAKENHI (Grant Number JP23K22796, JP23K19120, and JP25K01262), JST PRESTO (Grant Number JPMJPR245C), research grant from Murata Science and Education Foundation, MEXT Q-LEAP (Grant Number JPMXS0118067395), JST ASPIRE (Grant Number JPMJAP24C1), JST SPRING (Grant Number JPMJSP2110), and the Spintronics Research Network of Japan. This work was supported by "Advanced Research Infrastructure for Materials and Nanotechnology in Japan (ARIM)" of the Ministry of Education, Culture, Sports, Science and Technology (MEXT) (Proposal Number JPMXP12 24KT2474) for device fabrication.

**Author Declarations**

The authors declare no competing financial interest.

**Data Availability**

The data that support the findings of this study are available from the corresponding authors upon reasonable request.

**References**

(1) Robledo, L.; Bernien, H.; van Weperen, I.; Hanson, R. Control and Coherence of the Optical Transition of Single Nitrogen Vacancy Centers in Diamond. *Phys Rev Lett* **2010**, *105* (17), 177403.

(2) Siyushev, P.; Pinto, H.; Vörös, M.; Gali, A.; Jelezko, F.; Wrachtrup, J. Optically Controlled Switching of the Charge State of a Single Nitrogen-Vacancy Center in Diamond at Cryogenic Temperatures. *Phys Rev Lett* **2013**, *110* (16), 167402.

(3) Chen, D.; Mu, Z.; Zhou, Y.; Fröch, J. E.; Rasmit, A.; Diederichs, C.; Zheludev, N.; Aharonovich, I.; Gao, W. Optical Gating of Resonance Fluorescence from a Single Germanium Vacancy Color Center in Diamond. *Phys Rev Lett* **2019**, *123* (3), 033602.

(4) Bourgeois, E.; Jarmola, A.; Siyushev, P.; Gulka, M.; Hruby, J.; Jelezko, F.; Budker, D.; Nesladek, M. Photoelectric Detection of Electron Spin Resonance of Nitrogen-Vacancy Centres in Diamond. *Nat Commun* **2015**, *6* (1), 8577.

(5) Mizuochi, N.; Yamasaki, S.; Takizawa, H.; Morishita, N.; Ohshima, T.; Itoh, H.; Isoya, J. Continuous-Wave and Pulsed EPR Study of the Negatively Charged Silicon Vacancy with $S=3/2$ and $C_{3v}$ Symmetry in n-Type 4H-SiC. *Phys Rev B* **2002**, *66* (23), 235202.

(6) Widmann, M.; Lee, S.-Y.; Rendler, T.; Son, N. T.; Fedder, H.; Paik, S.; Yang, L.-P.; Zhao, N.; Yang, S.; Booker, I.; Denisenko, A.; Jamali, M.; Momenzadeh, S. A.; Gerhardt, I.; Ohshima, T.; Gali, A.; Janzén, E.; Wrachtrup, J. Coherent Control of Single Spins in Silicon Carbide at Room Temperature. *Nat Mater* **2015**, *14* (2), 164–168.

(7) Nagy, R.; Niethammer, M.; Widmann, M.; Chen, Y.-C.; Udvarhelyi, P.; Bonato, C.; Hassan, J. U.; Karhu, R.; Ivanov, I. G.; Son, N. T.; Maze, J. R.; Ohshima, T.; Soykal, Ö. O.; Gali, Á.; Lee, S.-Y.; Kaiser, F.; Wrachtrup, J. High-Fidelity Spin and Optical Control of Single Silicon-Vacancy Centres in Silicon Carbide. *Nat Commun* **2019**, *10* (1), 1954.

(8) Son, N. T.; Anderson, C. P.; Bourassa, A.; Miao, K. C.; Babin, C.; Widmann, M.; Niethammer, M.; Ul Hassan, J.; Morioka, N.; Ivanov, I. G.; Kaiser, F.; Wrachtrup, J.; Awschalom, D. D. Developing Silicon Carbide for Quantum Spintronics. *Appl Phys Lett* **2020**, *116* (19), 190501.

(9) Babin, C.; Stöhr, R.; Morioka, N.; Linkewitz, T.; Steidl, T.; Wörnle, R.; Liu, D.; Hesselmeier, E.; Vorobyov, V.; Denisenko, A.; Hentschel, M.; Gobert, C.; Berwian, P.; Astakhov, G. V.; Knolle, W.; Majety, S.; Saha, P.; Radulaski, M.; Son, N. T.; Ul-Hassan, J.; Kaiser, F.; Wrachtrup, J. Fabrication and Nanophotonic Waveguide Integration of Silicon Carbide Colour Centres with Preserved Spin-Optical Coherence. *Nat Mater* **2022**, *21* (1), 67–73.

(10) Carter, S. G.; Soykal, Ö. O.; Dev, P.; Economou, S. E.; Glaser, E. R. Spin Coherence and Echo Modulation of the Silicon Vacancy in 4H-SiC at Room Temperature. *Phys Rev B* **2015**, *92* (16), 161202.

(11) Parthasarathy, S. K.; Kallinger, B.; Kaiser, F.; Berwian, P.; Dasari, D. B. R.; Friedrich, J.; Nagy, R. Scalable Quantum Memory Nodes Using Nuclear Spins in Silicon Carbide. *Phys Rev Appl* **2023**, *19* (3), 034026.

(12) Udvarhelyi, P.; Thiering, G.; Morioka, N.; Babin, C.; Kaiser, F.; Lukin, D.; Ohshima, T.; Ul-Hassan, J.; Son, N. T.; Vučković, J.; Wrachtrup, J.; Gali, A. Vibronic States and Their Effect on the Temperature and Strain Dependence of Silicon-Vacancy Qubits in 4H-SiC. *Phys Rev Appl* **2020**, *13* (5), 054017.

(13) Morioka, N.; Babin, C.; Nagy, R.; Gediz, I.; Hesselmeier, E.; Liu, D.; Joliffe, M.; Niethammer, M.; Dasari, D.; Vorobyov, V.; Kolesov, R.; Stöhr, R.; Ul-Hassan, J.; Son, N. T.; Ohshima, T.; Udvarhelyi, P.; Thiering, G.; Gali, A.; Wrachtrup, J.; Kaiser, F. Spin-Controlled Generation of Indistinguishable and Distinguishable Photons from Silicon Vacancy Centres in Silicon Carbide. *Nat Commun* **2020**, *11* (1), 2516.

(14) Fang, R. Z.; Lai, X. Y.; Li, T.; Su, R. Z.; Lu, B. W.; Yang, C. W.; Liu, R. Z.; Qiao, Y. K.; Li, C.; He, Z. G.; Huang, J.; Li, H.; You, L. X.; Huo, Y. H.; Bao, X. H.; Pan, J. W. Experimental Generation of Spin-Photon Entanglement in Silicon Carbide. *Phys Rev Lett* **2024**, *132* (16), 160801.

(15) Hesselmeier, E.; Kuna, P.; Knolle, W.; Kaiser, F.; Son, N. T.; Ghezellou, M.; Ul-Hassan, J.; Vorobyov, V.; Wrachtrup, J. High-Fidelity Optical Readout of a Nuclear-Spin Qubit in Silicon Carbide. *Phys Rev Lett* **2024**, *132* (18), 180804.

(16) Lai, X.-Y.; Fang, R.-Z.; Li, T.; Su, R.-Z.; Huang, J.; Li, H.; You, L.-X.; Bao, X.-H.; Pan, J.-W. Single-Shot Readout of a Nuclear Spin in Silicon Carbide. *Phys Rev Lett* **2024**, *132* (18), 180803.

(17) Day, A. M.; Dietz, J. R.; Sutula, M.; Yeh, M.; Hu, E. L. Laser Writing of Spin Defects in Nanophotonic Cavities. *Nat Mater* **2023**, *22* (6), 696–702.

(18) Lukin, D. M.; Guidry, M. A.; Yang, J.; Ghezellou, M.; Deb Mishra, S.; Abe, H.; Ohshima, T.; Ul-Hassan, J.; Vučković, J. Two-Emitter Multimode Cavity Quantum Electrodynamics in Thin-Film Silicon Carbide Photonics. *Phys Rev X* **2023**, *13* (1), 011005.

(19) Körber, J.; Heiler, J.; Fuchs, P.; Flad, P.; Hesselmeier, E.; Kuna, P.; Ul-Hassan, J.; Knolle, W.; Becher, C.; Kaiser, F.; Wrachtrup, J. Fluorescence Enhancement of Single V2 Centers in a 4H-SiC Cavity Antenna. *Nano Lett* **2024**, *24* (30), 9289–9295.

(20) Anderson, C. P.; Bourassa, A.; Miao, K. C.; Wolfowicz, G.; Mintun, P. J.; Crook, A. L.; Abe, H.; Ul Hassan, J.; Son, N. T.; Ohshima, T.; Awschalom, D. D. Electrical and Optical Control of Single Spins Integrated in Scalable Semiconductor Devices. *Science* **2019**, *366* (6470), 1225–1230.

(21) Steidl, T.; Kuna, P.; Hesselmeier-Hüttmann, E.; Liu, D.; Stöhr, R.; Knolle, W.; Ghezellou, M.; Ul-Hassan, J.; Schober, M.; Bockstedte, M.; Bian, G.; Gali, A.; Vorobyov, V.; Wrachtrup, J. Single V2 Defect in 4H Silicon Carbide Schottky Diode at Low Temperature. *Nat Commun* **2025**, *16* (1), 4669.

(22) Scheller, D.; Hrunski, F.; Schwarberg, J. H.; Knolle, W.; Soykal, Ö. O.; Udvarhelyi, P.; Narang, P.; Weber, H. B.; Hollendonner, M.; Nagy, R. Quantum-Enhanced Electric Field Mapping within Semiconductor Devices. *Phys Rev Appl* **2025**, *24* (1), 014036.

(23) Niethammer, M.; Widmann, M.; Rendler, T.; Morioka, N.; Chen, Y.-C.; Stöhr, R.; Hassan, J. U.; Onoda, S.; Ohshima, T.; Lee, S.-Y.; Mukherjee, A.; Isoya, J.; Son, N. T.; Wrachtrup, J. Coherent Electrical Readout of Defect Spins in Silicon Carbide by Photo-Ionization at Ambient Conditions. *Nat Commun* **2019**, *10* (1), 5569.

(24) Nishikawa, T.; Morioka, N.; Abe, H.; Morishita, H.; Ohshima, T.; Mizuochi, N. Electrical Detection of Nuclear Spins via Silicon Vacancies in Silicon Carbide at Room Temperature. *Appl Phys Lett* **2022**, *121* (18), 184005.

(25) Nishikawa, T.; Morioka, N.; Abe, H.; Murata, K.; Okajima, K.; Ohshima, T.; Tsuchida, H.; Mizuochi, N. Coherent Photoelectrical Readout of Single Spins in Silicon Carbide at Room Temperature. *Nat Commun* **2025**, *16* (1), 3405.

(26) Manson, N. B.; Harrison, J. P. Photo-Ionization of the Nitrogen-Vacancy Center in Diamond. *Diam Relat Mater* **2005**, *14* (10), 1705–1710.

(27) Aslam, N.; Waldherr, G.; Neumann, P.; Jelezko, F.; Wrachtrup, J. Photo-Induced Ionization Dynamics of the Nitrogen Vacancy Defect in Diamond Investigated by Single-Shot Charge State Detection. *New J Phys* **2013**, *15* (1), 013064.

(28) Fuchs, F.; Stender, B.; Trupke, M.; Simin, D.; Pflaum, J.; Dyakonov, V.; Astakhov, G. V. Engineering Near-Infrared Single-Photon Emitters with Optically Active Spins in Ultrapure Silicon Carbide. *Nat Commun* **2015**, *6* (1), 7578.

(29) Todenhagen, L. M.; Brandt, M. S. Optical and Electrical Readout of Diamond NV Centers in Dependence of the Excitation Wavelength. *Appl Phys Lett* **2025**, *126* (19), 194003.

(30) Sze, S. M.; Coleman, D. J.; Loya, A. Current Transport in Metal-Semiconductor-Metal (MSM) Structures. *Solid State Electronics* **1971**, *14* (12), 1209–1218.

(31) Sze, S. M.; Ng, K. K. *Physics of Semiconductor Devices, 3rd Edition*; Wiley, 2007.

(32) Hain, T. C.; Fuchs, F.; Soltamov, V. A.; Baranov, P. G.; Astakhov, G. V.; Hertel, T.; Dyakonov, V. Excitation and Recombination Dynamics of Vacancy-Related Spin Centers in Silicon Carbide. *J Appl Phys* **2014**, *115* (13), 133508.

(33) Nagy, R.; Widmann, M.; Niethammer, M.; Dasari, D. B. R.; Gerhardt, I.; Soykal, Ö. O.; Radulaski, M.; Ohshima, T.; Vučković, J.; Son, N. T.; Ivanov, I. G.; Economou, S. E.; Bonato, C.; Lee, S.-Y.; Wrachtrup, J. Quantum Properties of Dichroic Silicon Vacancies in Silicon Carbide. *Phys Rev Appl* **2018**, *9* (3), 034022.

(34) Singh, H.; Hollberg, M. A.; Ghezellou, M.; Ul-Hassan, J.; Kaiser, F.; Suter, D. Characterization of Single Shallow Silicon-Vacancy Centers in 4H−SiC. *Phys Rev B* **2023**, *107* (13), 134117.

(35) Morioka, N.; Liu, D.; Soykal, Ö. O.; Gediz, I.; Babin, C.; Stöhr, R.; Ohshima, T.; Son, N. T.; Ul-Hassan, J.; Kaiser, F.; Wrachtrup, J. Spin-Optical Dynamics and Quantum Efficiency of a Single V1 Center in Silicon Carbide. *Phys Rev Appl* **2022**, *17* (5), 054005.

(36) Liu, D.; Kaiser, F.; Bushmakin, V.; Hesselmeier, E.; Steidl, T.; Ohshima, T.; Son, N. T.; Ul-Hassan, J.; Soykal, Ö. O.; Wrachtrup, J. The Silicon Vacancy Centers in SiC: Determination of Intrinsic Spin Dynamics for Integrated Quantum Photonics. *npj Quantum Inf* **2024**, *10* (1), 72.

(37) Booker, I. D.; Janzén, E.; Son, N. T.; Hassan, J.; Stenberg, P.; Sveinbjörnsson, E. Ö. Donor and Double-Donor Transitions of the Carbon Vacancy Related EH6/7 Deep Level in 4H-SiC. *J Appl Phys* **2016**, *119* (23), 235703.

(38) Son, N. T.; Ivanov, I. G. Charge State Control of the Silicon Vacancy and Divacancy in Silicon Carbide. *J Appl Phys* **2021**, *129* (21), 215702.

(39) Miyazawa, T.; Tsuchida, H. Point Defect Reduction and Carrier Lifetime Improvement of Si- and C-Face 4H-SiC Epilayers. *J Appl Phys* **2013**, *113* (8), 083714.

(40) Motoki, S.; Sato, S.; Saiki, S.; Masuyama, Y.; Yamazaki, Y.; Ohshima, T.; Murata, K.; Tsuchida, H.; Hijikata, Y. Optically Detected Magnetic Resonance of Silicon Vacancies in 4H-SiC at Elevated Temperatures toward Magnetic Sensing under Harsh Environments. *J Appl Phys* **2023**, *133* (15), 154402.

(41) Alfieri, G.; Monakhov, E. V.; Svensson, B. G.; Linnarsson, M. K. Annealing Behavior between Room Temperature and 2000 °C of Deep Level Defects in Electron-Irradiated n-Type 4H Silicon Carbide. *J Appl Phys* **2005**, *98* (4), 043518.

(42) Rühl, M.; Ott, C.; Götzinger, S.; Krieger, M.; Weber, H. B. Controlled Generation of Intrinsic Near-Infrared Color Centers in 4H-SiC via Proton Irradiation and Annealing. *Appl Phys Lett* **2018**, *113* (12), 122102.

(43) Storasta, L.; Tsuchida, H.; Miyazawa, T.; Ohshima, T. Enhanced Annealing of the Z1/2 Defect in 4H–SiC Epilayers. *J Appl Phys* **2008**, *103* (1), 013705.

(44) Hiyoshi, T.; Kimoto, T. Elimination of the Major Deep Levels in N- and p-Type 4H-SiC by Two-Step

Thermal Treatment. *Applied Physics Express* **2009**, *2* (9), 091101.

(45) Hiyoshi, T.; Kimoto, T. Reduction of Deep Levels and Improvement of Carrier Lifetime in N-Type 4H-SiC by Thermal Oxidation. *Applied Physics Express* **2009**, *2* (4), 041101.

(46) Karsthof, R.; Bathen, M. E.; Galeckas, A.; Vines, L. Conversion Pathways of Primary Defects by Annealing in Proton-Irradiated n-Type 4H-SiC. *Phys Rev B* **2020**, *102* (18), 184111.

(47) Xue, Y.; Titze, M.; Mack, J.; Yang, Z.; Zhang, L.; Su, S. S.; Zhang, Z.; Fan, L. Selective Generation of V2 Silicon Vacancy Centers in 4H-Silicon Carbide. *Nano Lett* **2024**, *24* (7), 2369–2375.

(48) Zappacosta, A.; Haylock, B.; Fisher, P.; Morioka, N.; Cernansky, R. *in preparation* **2025**.